\documentclass[aps,prc,nofootinbib,amsmath,amssymb]{revtex4}
\usepackage{graphicx}
\usepackage{hyperref}
\begin{document}

\title[]{Mass-Energy Equivalence in Bound Three-Nucleon Systems}

\author{I. Filikhin, V.M. Suslov and B. Vlahovic}

\affiliation{CREST, North Carolina Central University, 
Durham, NC 27707, USA}

\begin{abstract}
The mass defect formula reflects the equivalence of mass and energy for bound nuclear systems. We study three-nucleon systems $^3$H and $^3$He, considering the neutron and proton as indistinguishable particles ($AAA$ model) or taking into account the real masses of neutrons and protons ($AAB$ model).
We have focused on conceptual problems of the $AAA$ model, which is widely used for $3N$ calculations. In particular, the $AAA$ model is incompatible with the mass defect formula, which naturally corresponds to the $AAB$ model. In addition, the $AAA$ model has a cyclic permutation symmetry, which is breaking in the natural $AAB$ model. The latter problem cannot be eliminated within the perturbative $AAA$ approach, in which the mass difference effect is simulated by correcting the kinetic energy operator.
   Earlier it was reported that the accuracy of such $AAA$ calculations is 1~keV.
An example of the $AAB$ calculation, we numerically estimate the effect of the difference between the neutron and proton masses on the energy calculated without any approximation with the accuracy of 0.1~keV.
Another manifestation of the equivalence of mass $m$ and energy $E$ can be expressed by the formula $dE/dm=Const$. To show this dependence of the three-body energy on the nucleon mass, we performed realistic calculations within the $AAA$ approximation, varying the averaged nucleon mass. The mass-energy compensation effect for the three-body Hamiltonian is shown. According to this, we have determined the effective nucleon mass required to compensate for the perturbative effect of a three-body potential.
\end{abstract}


\maketitle

\section{Introduction}
Within the isospin formalism, the neutron and proton are considered identical particles. For three-nucleon systems, such simplification can be described as an $AAA$ model in which the
proton-neutron ($p$-$n$) mass difference is ignored. Traditionally, the neutron-proton mass difference is ignored in the realistic 3$N$ calculations \cite{DS}
because the effect of this difference is estimated as small \cite{Nogga1}.
However, this CD effect for the deiffrnece of $^3$H and $^3$He bound energies, evaluated in early works \cite{Br88,A} is significant and amounts to 12~keV.
Recently, we proposed an $AAB$ model,  where the $n$-$p$ mass difference and the charge dependence of nucleon-nucleon potential were taken into account \cite{FSV16a,FSV16,FSV19}. The results of Ref. \cite{FSV16} for the mass difference effect for $^3$H is 2~keV, and for $^3$He is 7~keV calculated using a semi-realistic potential. Similar calculations \cite{NM18}  performed with different $NN$ potentials resulted in the differences from 4~keV to 7~keV for the $^3$H.
Comparing both models, $AAA$ and $AAB$, we assume that the mass difference adds uncertainty of several keV to $AAA$ calculations for the $^3$H binding energy when a comparison with experimental data is proposed. This uncertainty is above the numerical accuracy of $\leq 1$ keV evaluated for the realistic $AAA$ calculations in Refs. \cite{N,R}. In addition, it should be noted that the numerical $AAA$ estimates of the energy correction motivated by the mass difference, obtained in the earlier calculation \cite{Nogga1} and in the recent \cite{M21}, are the same when using different potentials.

In the present work, we discuss the $AAA$ calculations in relation to the mass defect formula aiming to show that rigorous treatment of the three-nucleon systems requires utilization of the $AAB$ model. The mass defect formula represents the relation between mass and energy for bound nuclear systems and reflects fundamental mass-energy equivalence. One can expect that realistic three-body calculations will follow this relation. One can find examples of such calculations in Refs.
 \cite{KKF,C85,Nogga,KVR} where the charge-independent Argonne AV14 potential \cite{AV14} was used. More rigorous treatment must include the charge dependence, which is detected by experimental evidence of a difference of spin-singlet scattering lengths for $nn$, $np$, and $pp$ interactions \cite{MO,Chen}.
In particular, the Argonne AV18 model for the $NN$ interaction takes into account the violation of the charge independence \cite{W2a}. As an example, the AV18 potential was applied for the Faddeev calculations in configuration space in Ref. \cite{LC,L}. The chiral potentials were used in recent calculations \cite{M,Sk,PT} with taking into account the charge symmetry independence breaking.
These calculations were performed within the $AAA$ model that ignores the $p$-$n$ mass deference. 

The realistic calculations are  associated with using a three-nucleon force (3NF).
The reason is that the experimental data for the $3N$ binding energy are not reproduced with realistic pair potentials. Three-nucleon force has been proposed \cite{P01,P08} as a solution to the problem (see Refs. \cite{Nogga1,ep,G,GKV,Yang21,V2022} and the review \cite{PT}).
To discuss another manifestation of the mass-energy equivalence in three-nucleon systems, we introduce an effective nucleon mass  within the $AAA$ model and show
that a decrease of a kinetic-energy term can be considered as compensation for the attractive contribution of three-body force.
The corresponding calculations for $^3$H and $^3$He nuclei are performed utilizing the Faddeev method in configuration space. We apply the  Argonne AV14 potential to compare it with the results of other calculations available in the literature with the same potential. The estimations for the kinetic and potential terms of the Hamiltonian are compared with calculations of different authors \cite{KKF,C85,Nogga,KVR} for $^3$H nucleus using different 3NFs. 
The competition between the kinetic and potential terms in the  Schr\"{o}dinger equation is well known. Such competition, for example, explains the effect arising from the dimension decrease in the equations describing charged trions ($eeh$ end $hhe$ systems) \cite{RK}.

Note that the idea of effective mass is widely used in solid-state physics \cite{K}. For example, \cite{FSV06}, the effective mass electron in InAs is 0.024$m_0$, whereas the mass in GaAs is 0.064$m_0$, where $m_0$ is the free electron mass. In nuclear and astrophysics, the effective mass approach takes place in the study of neutron stars \cite{em} within a more general single-particle method of many-body physics \cite{C}. According to the consideration, a single particle is in a complex interaction with a medium \cite{La,B} and the single-particle potential is assumed to be energy dependent \cite{MB}.
Our work is the first proposition for a phenomenological approach based on the effective mass idea to describe theoretical and experimental results for 3$N$ systems. In particular, we provide a mechanism of the compensation for three-body force following the most general definition for a particle's effective mass as the mass that seems to have when responding to the force.

\section{Definitions}
The starting point for studying three-nucleon systems is the solution of the Schr\"odinger equation
for nuclear Hamiltonian:
\begin{equation}
\label{H}
 H =
-\frac{\hbar^2}{2m}\sum_{i=1}^3\nabla^2_i + \sum_{j<k} V_{jk}+V_{3N},
\end{equation}
where $V_{jk}$ and
$V_{3N}$ are the two- and three-body nuclear potentials, respectively. Here, the  index  $i$ ($j$ and $k$) numerates the particles, $i=1,2,3$, $m$ is averaged mass of nucleons.
In the presented work, we neglect the three-nucleon forces, $V_{3N}$=0. 
The  Schr\"odinger equation is equivalent to the set of the Faddeev equations for components of the total wave function. 
The general form of the  Faddeev equations in coordinate space
  is as follows \cite{FM}:
\begin{equation}
\{H_0^\alpha+V^s_{\alpha}(\vert\vec{x}_{\alpha}\vert)+
\sum_{\beta =1}^3V^{Coul.}_{\beta}(\vert\vec{x}_{\beta}\vert)-E\}
\Psi_{\alpha}(\vec{x}_{\alpha},\vec{y}_{\alpha})
=-V^s_{\alpha}(\vert\vec{x}_{\alpha}\vert)
\sum_{\beta\ne\alpha}\Psi_{\beta}(\vec{x}_{\beta},\vec{y}_{\beta}),
\label{F0}
\end{equation}
where $V^{Coul.}_{\beta}$ is the Coulomb potential between the particles
belonging to the pair $\beta$ and $V^s_\alpha$ is the short-range
pair potential in the two-body channel $\alpha$, ($\alpha$=1,2,3).
$H_0^\alpha=-\Delta_{\vec{x}_{\alpha}}-\Delta_{\vec{y}_{\alpha}}$ is the
kinetic energy operator with $\hbar^2=1$, $E$ is the total energy, and $\Psi$ is the wave
function of the three-body system. $\Psi$ is given as a sum over
three Faddeev components, $\Psi =\sum^3_{\alpha=1}\Psi_\alpha$.  Here, 
$(\vec{x}_{\alpha} ,\vec{y}_{\alpha})$ are  the mass-scaled Jacobi coordinates. 
The Jacobi vectors with
different $\alpha$'s are linearly related by the orthogonal transformation
\begin{equation}
\label{C0}
  \left(
  \begin{array}{c}
     \vec{x}_{\alpha} \\ \vec{y}_{\alpha}
  \end{array}
  \right)=
  \left(
  \begin{array}{rl}
      C_{\alpha\beta} & S_{\alpha\beta} \\
     -S_{\alpha\beta} & C_{\alpha\beta}
  \end{array}
  \right)
  \left(
  \begin{array}{c}
     \vec{x}_{\beta} \\ \vec{y}_{\beta}
  \end{array}
  \right) \ ,\ \ \ C^2_{\alpha\beta} + S^2_{\alpha\beta} = 1,
\end{equation}
where
\begin{equation}
\label{C1}
\begin{array}{c}
C_{\alpha\beta}=-\sqrt{\frac{m_{\alpha}m_{\beta}}
{(M-m_{\alpha})(M-m_{\beta})}}, \ \
S_{\alpha\beta} = (-)^{\beta - \alpha}{\rm sgn}(\beta - \alpha)
\sqrt{1-C^{2}_{\alpha\beta}}, \\
M=\sum_{\alpha=1}^3m_{\alpha}.
\end{array}
\end{equation}

\subsection{\texorpdfstring{$AAA$}\ \ {model}}
Within the isotopic spin formalism,  the particles of the three-nucleon system are considered identical. The mass difference of the nucleons is ignored in the $AAA$ model.
 The set of the Faddeev equations (\ref{F0}) can be reduced 
by a single equation of the form \cite{FM}:
\begin{equation}
(H_{0}+V_{AA}-E)U=-V_{AA}(P_c^++P_c^-)U,
\label{EQ}
\end{equation}%
where $V_{AA}$ is a short-range $AA$ potential.
$H_0=-\frac{\hbar ^{2}}{m}(\Delta_{\bf x}+\Delta_{{\bf y}})$ is the internal  kinetic energy operator,  and $m$  is the  averaged nucleon mass. 
The total wave function has a cyclic permutation symmetry expressed as 
\begin{equation}
\Psi=(I+P_c^++P_c^-)U, 
\label{EQ0}
\end{equation}
where $P_c^{+}$ and $P_c^{-}$ are the nucleon cyclic permutation operators.
Jacobi vectors ${\vec{x}_\alpha , \vec{y}_\alpha }$ can be taken as independent coordinates.
For the $\alpha $=1 pair, they are related to the particle coordinates by the formulas:
\begin{equation}
\label{x}
\vec{x}_1=\vec{r}_2-\vec{r}_3, \ \ \ \ \ \ \vec{y}_1 =%
\frac{\vec{r}_2+\vec{r}_3}2 -\vec{r}_1,
\end{equation}
for $\alpha $=2,3, one has to make cyclic permutations of the indexes in Eq. (\ref{x}). 
Here, $\vec{r}_\alpha$ are coordinates of the particles relative to the center of mass of the three-body system. 
The  cyclic permutation operators in Eq. ({\ref{EQ}) transform the coordinates associated with the  pair $\alpha $
to the pair $\beta$.
According to Eqs. (\ref{C0}) and (\ref{C1}), the coordinate transformation for an $AAA$ system has the simplest form as it is also seen by Eq. (\ref{x}),
where the transformation does not depend on the mass.
For systems including particles of different masses, the transformation of the Jacobi coordinates will include the masses of particles according to Eqs. (\ref{C0})-(\ref{C1}) (for instant see Ref. \cite{LC19}). 

\subsection{\texorpdfstring{$AAB$}\ \ model}
For a three-body system with two identical particles (the $AAB$ model) the
set of the Faddeev equations is presented by
two equations for the components $U$ and $W$ \cite{K91,MG,F18}:
\begin{equation}
\begin{array}{l}
{(H^U_{0}+V_{AA}-E)U=-V_{AA}(W\pm PW),} \\
{(H^W_{0}+V_{AB}-E)W=-V_{AB}(U \pm PW),}
\end{array}
\label{EQ1}
\end{equation}%
where  the signs ''$+ $'' and ''$- $'' correspond to two identical bosons and fermions, respectively and $ H^{U}_{0}$ and $ H^{W}_{0}$
are the kinetic energy operators presented in the Jacobi coordinates for corresponding particle rearrangement and channel masses.
The total wave function of the $AAB$ system is decomposed into the sum of the Faddeev components $U$ and $W$ corresponding to the $(AA)B$ and $A(AB)$ types of rearrangements: 
\begin{equation}
\Psi =U+W\pm PW, 
\label{EQ2}
\end{equation}
where $P$ is the permutation operator for two identical particles. The wave function of the system $AAB$ becomes to be antisymmetrized for two identical fermions automatically.

Comparing Eqs. (\ref{EQ2}) and (\ref{EQ0}), one can see that the wave function $\Psi $ given by Eq. (\ref{EQ2}) loses the symmetry of cyclical permutations of Eq. (\ref{EQ0}) due to forbidding a permutation of the non-identical particles $A$ and $B$.

\subsection{Mass correction for \texorpdfstring{$AAA$} \ \ model}
The physical masses of particles in the 3$N$ system ($nnp$ or $ppn$) are different like in an $AAB$ system. Thus, the direct use of the $AAA$ model is possible only in an approximation that does not take into account the slight difference in the masses of neutrons and protons.
The correction for the $AAA$ model should mimic the properties of the corresponding $AAB$ systems. This modeling has two aspects. The first is related to the transformations of the Jacobi coordinates corresponding to the permutation operators in the equations for the Faddeev components. Coordinate transformations differ for the $AAA$ and $AAB$ systems according to the equation (\ref{C0}). The $AAA$ approximation cannot mimic this difference. In the case of the assumption about the nucleon mass difference,  Eq. (\ref{EQ}) has to be replaced by Eq.  \ref{EQ1}. The wave function of a $AAB$ system loses the cyclical permutation symmetry which is attributed to a $AAA$ model.

The second aspect concerns the modeling of the  $AAB$ kinetic energy operator $H_0$.
A correction for the binding energy 3$N$ for the AAA model was proposed in \cite{F90} as a correction to the kinetic operator $H_0$: $H_0+\Delta H_0$, where $\Delta H_0$ includes
isospin variables (similar to the Coulomb potential) and the mass difference $\Delta m=m_n-m_p$.
 The mass of the $i$-th nucleon can be expressed as \cite{F90}:
\begin{equation}
\label{m}
m_i =m - \Delta m \tau_z^i/2,
\end{equation}
where $\Delta m=m_n-m_p$.
The matrix element of $\Delta H_0$ can be evaluated in the first order of the perturbation theory \cite{F90} like
$$ \langle \Delta H_{0} \rangle =  \langle \Psi \vert \frac{\Delta m}{2m}\sum_{i=1,2,3}\frac{ \pi_i^2}{2m}\tau_z^i\vert \Psi \rangle.$$
This term  is related to  the charge symmetry breaking by the mass difference and contributes to the energy difference of $^3$He and
$^3$H as $2\langle \Delta H_{0} \rangle $\cite{F90} (or more complicated with the Coulomb correction in Ref. \cite{M21}).
The corresponding perturbative correction is also possible for the wave function. However, this correction was practically not applied to the 3$N$ calculations. Note that such correction of the wavefunction does not change the cyclic permutation symmetry associated with the $AAA$ model.

The correction for mass can also be made for bosonic systems. In this case, the particles must be numbered manually to model the expression (\ref{m}).
In this regard, a slightly different procedure for making the correction can be proposed to simulate the $AAB$ system within the $AAA$ model. The particles of the system have masses $m_A$, $m_A$, and $m_B$. We can number the masses as
$$
\begin{array}{l}
m_1=(m_A+m_B)/2+ m_A/2-m_B/2= m+\Delta m/2,\\
m_2=(m_A+m_B)/2+ m_A/2-m_B/2= m+\Delta m/2,\\
m_3=(m_A+m_B)/2+ m_B/2-m_A/2= m-\Delta m/2,
\end{array}
$$
where $\Delta m=m_A-m_B$ is small. Identical particles $A$ can be particle 1 and particle 2.
Following  Ref. \cite{F90}, we write the kinetic energy operator  in terms of the individual momenta of the particles in the  center-of-mass frame:
\begin{equation}
\label{cm}
\hat{H_0}= \sum_{i=1,2,3}\frac{ \pi_i^2}{2m_i}\approx \sum_{i=1,2,3}\frac{ \pi_i^2}{2m}-\frac{\Delta m}{2m}\frac{ \pi_1^2}{2m}-\frac{\Delta m}{2m}\frac{ \pi_2^2}{2m}+\frac{\Delta m}{2m}\frac{ \pi_3^2}{2m}.
\end{equation}

In the first order perturbation theory, the correction for energy can be presented as  $\langle {\hat{H_0}} \rangle \approx \langle H_0 \rangle+\langle \Delta H_0\rangle$, with the matrix element
\begin{equation}
\label{D}
\langle \Delta H_{0} \rangle = \langle \Psi^R(\vec{r_1},\vec{r_2},\vec{r_3})\vert -\frac{\Delta m}{2m}\frac{ \pi_1^2}{2m}-\frac{\Delta m}{2m}\frac{ \pi_2^2}{2m}+\frac{\Delta m}{2m}\frac{ \pi_3^2}{2m} \vert \Psi^R(\vec{r_1},\vec{r_2},\vec{r_3})\rangle,
\end{equation}
where $\Psi^R(\vec{r_1},\vec{r_2},\vec{r_3})$ is the coordinate part of the wave function $\Psi=\xi_{spin}\otimes \eta_{isospin}\otimes \Psi^R$ obtained within an $AAA$ model.
In this approach, the mass correction for the $AAA$ energy is isospin-independent.
One can interpret the $\Delta H_0$ term as a part of a three-body force that acts in all partial waves.
This example shows the mass isospin-dependence of the correction proposed in Refs. \cite{F90} is artificial and is not practically needed for the systems mentioned.  
The isospin(spin) basis for the  $AAA$ and $AAB$ systems is the same. 
However, the scalar variant of a perturbative mass correction for the $AAA$ model is still an approximation based on the small value of the mass difference.

Finally, we would like to note that the correction (\ref{D}) of the $AAA$ Hamiltonian makes a possibility to define an ''corrected particle mass''. Based on the expression (\ref{D}) for the term $\Delta H_0$, one can rewrite Eq. (\ref{cm}) like 
\begin{equation}
\hat{H_0}= \sum_{i=1,2,3}\frac{ \pi_i^2}{2m_i}\approx \sum_{i=1,2,3}\frac{ \pi_i^2}{2m'},
\end{equation}
where $m' = m+\delta m$. The $\delta m$ can be adjusted to simulate the contribution $\langle \Delta H_{0} \rangle$ to the binding energy.

Taking into account the results of the $^3$H calculations given in Ref. \cite{M21}, one can roughly estimate the value of the correction for the nucleon mass using the relation 
\begin{equation}
\delta m/m \approx 2\langle \Delta H_{0} \rangle / \langle H_{0} \rangle. 
\label{mm}
\end{equation}
This estimate gives a value of $\delta m$ about 0.3~MeV with N2LO+3NFs  potentials due to the $\langle \Delta H_{0} \rangle$ has been estimated about 6~keV and  the averaged kinetic energy has been calculated to be about 35~MeV. The next step could be a new calculation for these matrix elements. In the new calculation, the corrected mass (+0.3 MeV) obtained in the first step is used. Ideally, in this iterative process based on Eq. (\ref{mm}), one can obtain $\langle \Delta H_{0} \rangle \approx 0$ and final values of the corrected mass and energy. It will be averaded nucleon mass which  exactly corresponds to the $AAA$ model.

The smallness of $\Delta m$ provides good accuracy in such  $AAA$ calculations for the 3$N$ binding energy \cite{M21}. 
For the hypernuclear  system $\Lambda NN$, the mass difference $\Delta m_\Lambda=m_\Lambda-m_N$ is about 176.12~MeV. This value is too large to use the $AAA$ model. According to this, in Ref. \cite{MG}, the $AAB$ model given by Eq. (\ref{EQ1}) has been used. 

It can be noted that the averaging of nucleon masses is appropriate for the system $\Lambda np$ when a charge independent $ \Lambda N$  potential is applied. After correction to physical masses for nucleons, we obtain an $ABC$ system, where $A$ and $B$ are nucleons and $C$ is the hyperon. The binding energy does not depend on the neutron/proton mass correction due to a compensation of kinetic energy terms related to the nucleons.
 The mass correction is expressed by the relation:
 $$
\begin{array}{l}
m_1=(m_A+m_B)/2+ m_A/2-m_B/2= m+\Delta m/2,\\
m_2=(m_A+m_B)/2+ m_B/2-m_A/2= m-\Delta m/2,\\
m_3=m_C,
\end{array}
$$
 where $m_3$ is the mass of the $\Lambda$ hyperon, $m$ is the averaged nucleon mass.
 The energy correction (\ref{D}) could be written in the form:
 \begin{equation}
\label{D1}
\langle \Delta H_{0} \rangle = \langle \Psi^R(\vec{r_1},\vec{r_2},\vec{r_3})\vert -\frac{\Delta m}{2m}\frac{ \pi_1^2}{2m}+\frac{\Delta m}{2m}\frac{ \pi_2^2}{2m}\vert \Psi^R(\vec{r_1},\vec{r_2},\vec{r_3})\rangle,
\end{equation}
The matrix element (\ref{D1}) is equal to zero in the first order of perturbation theory. Numerical calculations \cite{Ke} for the $KK\bar K$ system confirm this effect. The $AAB$ model for the $KK\bar K$ system was also studied in Ref. \cite{D}. The effect of the difference in the masses of kaons and antikaon is clearly visible for the binding energy and is expressed in values up to one MeV, in contrast to $3N$ systems where it is of the order of several keV. The reason for the difference is that the relative values of corrections for the proton-neutron and kaon-antikaon masses $\Delta m/m$ for the systems differ by a factor of 4.

\subsection{Mass-energy compensation}
We can make a transformation for the Hamiltonian  (\ref{H}) of the AAA model written as a linear combination
 $\beta H_0+\alpha V$, 
where $\beta, \alpha> 0$, and $V$ includes pair potentials. The parameters $\beta$ and $ \alpha$
are considered in a small vicinity of the point $\beta = \alpha =1$.
The calculated binding energy $E$ becomes a function of the $\alpha$ and $\beta$: $E=E(\alpha,\beta)$.
Upon averaging, one has:
\begin{equation}
\beta\langle H_0 \rangle+\alpha\langle V\rangle=E(\alpha,\beta).
\label{scal}
\end{equation}
As it is well known, the terms $\langle H_0 \rangle$ and $\langle V\rangle$ have 
opposite signs. Thus, small changes of the potential term given as  $\alpha$$\langle V\rangle$ can be compensated by changings of the  $\beta$. We definite the property as the mass-energy compensation.
As an illustration, we propose a nucleon effective mass
for three-nucleon systems. Below, will be shown that the perturbative contribution of three-body force 
may be compensated by variations of the averaged nucleon mass.

Considering small variations for the parameters $\alpha$ and $\beta=m/m^*$, where $\alpha=1+a$, $m^*=m+\Delta m$ and $|a|, |\Delta m/m|<1$ one can definite an effective mass   $m^*$.
The corresponding variations for the left-hand side of Eq. (\ref{scal}) can be written as
$$
\langle \Delta H_0 \rangle\approx b\Delta m, \quad  \langle \Delta V\rangle =a\langle V\rangle .
$$
Here, we use the Tailor's expansion of the function $1/m^*$ around the point $m$  taking the first two terms of the expansion like in Eq. (\ref{cm}). Such approximation leads to the linear dependence of the energy correction on $\Delta m$ in Eq.~(\ref{scal}). The coefficient $b$ is defined as $\frac1{4m^2}\langle \Psi^R(\vec{r_1},\vec{r_2},\vec{r_3})\vert - {\pi_1^2}-{ \pi_2^2}+{ \pi_3^2} \vert \Psi^R(\vec{r_1},\vec{r_2},\vec{r_3})\rangle$.
 
\section{ The mass defect formula and \texorpdfstring{$AAA$} \ \ calculations}
\label{sec:md}
According to the definition of the mass defect, the energy of the $^3$H ground state $E_t$ can be calculated as the following:
\begin{equation}
\label{dm}
E_t=M_t-2m_n-m_p,
\end{equation}%
where $m_n$=939.56542052 MeV, $m_p$=938.27208816 \cite{data} are free neutron and proton masses at rest. 
$M_t$=2808.92113298 MeV is the triton mass \cite{data}.
One can obtain: $E_t=M_t-2m_n-m_p$=-8.48179622 MeV that is close to the value of -8.481798 $\pm$ 0.000002 MeV reported in Ref. \cite{tunl1}.

We assume that
within the $AAB$ model, three-body calculations for the $^3$H energy $E_t^{Cal.}(m_n,m_p)$ can be considered as a function of the input parameters $m_n$ and $m_p$.
According to Eq. (\ref{dm}), we have
\begin{equation}
\label{df00}
E_t^{Cal.}(m_n,m_p)=M_t-2m_n-m_p.
\end{equation}
For the $AAA$ model, the nucleon has the mass $m_N$, which is calculated as an averaged mass of neutron and proton. The results of the three-body calculation $E_t^{Cal.}(m_N,m_N)$ is a function of the averaged nucleon mass.

Formally, the following relation has to be satisfied:
\begin{equation}
\label{df0}
E_t^{Cal.}(m_N,m_N)=M_t-3m_N.
\end{equation}
There is common agreement that the approximation $$E_t^{Cal.}(m_N,m_N)\approx E_t^{Cal.}(m_n,m_p)$$ is appropriate \cite{K2008}, because the $^3$H binding energy weakly depends on the proton and neutron mass difference.
However, the right-hand sites of the Eqs. (\ref{df00}) and (\ref{df0}) are different.
One can write the sum of the particle masses corresponding to the right-hand side of Eq.  (\ref{df0}) as 3$m_N=2m_n+m_p+(m_p-m_n)/2$ (or 3$m_N=2m_p+m_n+(m_n-m_p)/2$).  Thus, the total mass of particles of the $AAA$ system $M(AAA)=3m_N$ can be expressed by
the sum $M(AAB)=2m_n+m_p$ of the physical masses:
\begin{equation}
M(AAA)=M(AAB)+(m_p-m_n)/2.
\end{equation}
The difference between the total mass of particles of  the $AAA$ model and the total physical masses
is about $\pm$0.647~MeV, where '$-$' is for $^3$H (and '$+$' is for $^3$He). 
This amount is comparable with the Coulomb energy shift of $^3$He relatively $^3$H \cite{K2008}. The disagreement between the $AAA$ model and the mass defect formula is visible clearly.  For the $^3$H ($^3$He) nucleus,  the sum of masses in the $AAA$ model is smaller (larger) than the sum in the physical $AAB$ model. We consider this disagreement as a methodological problem: the $AAA$ model remains a non-rigorous approximation until the difference in masses is not taken into account.
To avoid the problem, one may calculate the energy of the three-nucleon system using the $AAB$ model. An example of such calculation is presented in Section~\ref{sec:AAB}.

Another way to improve the $AAA$ approximation was given in Ref. \cite{K2008} where the mass difference was taken in a perturbative manner using an matrix element of the operator of the kinetic energy written as
$K_\Delta =\sum_{i=1,3}\frac12(\frac1{2m_p}-\frac1{2m_n})\nabla^2_i\tau_z(i)$, were $\tau_z(i)$ is the isospin projection operator of $i$-th particle.
The corrected value for $E_t^{Cal.}(m_N,m_N)$ becomes closer to $E_t^{Cal.}(m_n,m_p)$.
Formally, this approach allows for avoiding the principal problem related to the mass defect formula due to the input parameters of this approach including the physical masses of nucleons. However, the approach keeps the $AAA$ model as a core approximation. In particular, the Jacobi coordinate sets which must correspond to the Faddeev components in  Eq. (\ref{EQ1}) are used without the mass difference. In another word, the cyclic permutation symmetry of the Faddeev components in Eq. (\ref{EQ}) is broken in the case of non-identical masses of particles.  One has to use  Eq. (\ref{EQ1}) to take into account new permutation symmetry.
Also,  this $AAA$-like model considers the pair potentials for the $nn$, $pp$, and $np$ pairs with the same averaged mass $m_N$.On the opposite, the $AAB$ model requests a definition of different averaged masses for these pairs of particles. As a result, the two-body potentials for nn and pp pairs must be corrected.  

A similar approximation has been proposed in Ref. \cite{W13} to estimate the effect of the mass-difference on the energy shift of $^3$He relatively $^3$H, $E(^3$He$)-E(^3$H$)$. However, the effect on the energy of each nucleus like $E_t^{Cal.}(m_n,m_p)-E_t^{Cal.}(m_N,m_N)$ cannot be separatelly defined within such approach. The $AAA$ results were calculated as averaged value $(E(^3$He$)-E(^3$H$))/2$ to approximate of $E_t^{Cal.}(m_n,m_p)-E_t^{Cal.}(m_N,m_N)$ and
$E_h^{Cal.}(m_n,m_p)-E_h^{Cal.}(m_N,m_N)$ which are corresponding to the mass-difference shifts in $^3$H and $^3$He. Thus, this $AAA$ model results in the same value for the two different cases \cite{Nogga1}. At the same time, our $AAB$ calculations \cite{FSV16a} for the mass-difference effect led to asymmetric values of  3~keV and 7~ keV for $^3$H and $^3$He, respectively, which is significantly inconsistent with the simplified estimation of the $AAA$ model.
More precise estimates Ref. \cite{M21} with the $AAA$ model lead to the values 5.840~keV and 5.26~keV respectively. Hovewer, this is also inconsistent with the results of the work \cite{FSV16a}.

\section{Effect of proton-neutron mass difference within \texorpdfstring{$S$}\ \ -wave \texorpdfstring{$AAB$}\ \ model}
\label{sec:AAB}
The $s$-wave $AAB$ model for bound states of three-nucleon systems was proposed in Ref. \cite{FSV16a,FSV16,FSV19} based on the "charge isospin basis" \cite{n2}.
The basis allows us to ''separate'' isospin variables. The charge isospin basis can be obtained by a unitary transformation of the ''natural'' isospin basis. The charge basis is widely used in the kaonic cluster study ($NN\bar K$ and $\bar K\bar K N$). Details of the consideration and references one can find in Ref. \cite{n2}.

We can note the interesting property of the potential model used in \cite{FSV16a,FSV16,FSV19}. The unitary transformation of the corresponding Hamiltonian leads to the final equations which can be described as “isospin-less” due to the same set of equations can be obtained without the assumption about the existence of isospin variables. Mainly, it is based on the property of the operators of pair interactions used in the $s$-wave model.  Before, this property has been mentioned in the literature \cite{SDG}.  The results of 3$N$ calculations do not depend on what the isospin basis is applied in formalism. 

Using the $AAB$ model, we evaluated the charge symmetry-breaking effect for the $^3$H and $^3$He nuclei. Based on the semi-realistic MT-I-III potential \cite{Fr}
defined by $np$ binding and scattering parameters, we have made a modification to MT-I-III(M) that reproduces the experimental data for
spin-singlet scattering lengths for $nn$, $np$, and $pp$ interactions. After that,
the differential Faddeev equations for the $nnp$ system are formulated as follows:
\begin{equation}
\label{eq:3}
\begin{array}{l}
(H_0^U+V_{nn}-E){\cal U}=-V_{nn}A(1+p){\cal W} \;\; , \\
(H_0^W+V_{np}-E){\cal W}=-V_{np} (A^T{\cal U}+Bp{\cal W}) \;\; ,
\end{array}
\end{equation}
where the exchange operator $p$ acts on coordinates only, $V_{nn}=v^s_{nn}$ is the spin singlet $nn$
potential, $V_{np}=diag\{v^s_{np},v^t_{np}\}$ is a diagonal $2\times2$ matrix with $v^s_{np}$ and $v^t_{np}$
the spin singlet and triplet $np$ potentials respectively,
and
\begin{equation}
\label{eq:matrix}
A=(-\frac12,\frac{\sqrt3}2), \qquad B=\left( \begin{array}{rr}-\frac12& -\frac{\sqrt3}2 \\-\frac{\sqrt3}2& \frac12 \\
\end{array} \right), \qquad
{\cal W}=\left( \begin{array}{rr}
{\cal W}_{1} \\
{\cal W}_{2} \\
\end{array} \right).
\end{equation}

To test the accuracy of our calculations, we add the result of the benchmark from Ref. \cite{K2008} in Tabl. \ref{t33} and calculated the $^3$H ground state energy with the parameters
of the MT-I-III potential given in Ref. \cite{K2008} and with the averaged nucleon mass $m_N$=938.94~MeV corresponding to $\hbar ^2/m_N$ = 41.47~MeVfm$^2$.
Both results coincide with four digits after the decimal point.
The results of our calculations for the binding energy of the $^3$H nucleus are presented in Tabl. \ref{t33}. One can compare the results for $AAA$ and $AAB$ models for different conditions.
In particular, the $AAB$ model makes it possible to directly estimate the effect of the nucleon mass difference on the binding energy $^3$H. The impact was estimated as 
\begin{equation}
\Delta E_m=E_t^{Cal.}(m_N,m_N)-E_t^{Cal.}(m_n,m_p).
\label{dm1}
\end{equation}
The corresponding values can be found in Tabl. \ref{t33}. The calculation gives  $\Delta E_m$=4.7~keV which agrees with the results of Ref. \cite{NM18}.
\begin{table}[!ht]
\caption{
The $^3$H ground state energies for MT-I-III and MT-I-III(M) potential within $AAA$ and $AAB$ models with averaged nucleon mass ($ m_n=m_p=m_N$),
with the physical masses of nucleons ($ m_n\neq m_p$) and with the charge dependence (CD). The contribution of each term of the Hamiltonian is shown in MeV. Here, $\hbar ^2/m_N$ = 41.471 MeV fm$^2$, $m_N$=938.919 MeV. The result of the calculation from Ref. \cite{K2008} is shown in parentheses. The result of realistic calculations from Ref. \cite{M21} is obtained with NLO potential within the mass corrected $AAA$ model ($AAA_{corr.}$).
}
\label{t33}
{\begin{tabular}{llccccc} \hline\noalign{\smallskip}
Model& Potential & $E$($^3$H) & $\Delta E_m$($^3$H)&$\langle H_0 \rangle$ & $\langle V_{2bf}\rangle$ & $\langle H_0+ V_{2bf} \rangle$ \\ \noalign{\smallskip}
\hline\noalign{\smallskip}
$AAA$ ($ m_n$=$m_p$=$m_N$) &MT-I-III & -8.5355 &-- &30.1 & -38.6 & -8.5 \\
$AAA$ ($ m_n$=$m_p$=$m_N$) &MT-I-III\cite{K2008}$^*$&-8.5357 & -- & -- & --  & --\\
 & &  (-8.5357) & -- &--  &-- & -- \\
$AAB$ ($ m_n\neq m_p$) & MT-I-III & -8.5402  & 4.7 & 30.1 & -38.6 & -8.5 \\
$AAB$ ($ m_n\neq m_p$, CD) & MT-I-III(M)& -8.3797  &-- &29.8 & -38.2 & -8.4 \\ 
$AAA_{corr.}$ ($ m_n\neq m_p$, CD) \cite{M21}& NLO& -8.325&6 & 35.87  &  -44.19 & -8.332 \\
\noalign{\smallskip}\hline
\end{tabular} \\
{\smallskip}
$^*$ $\hbar^2/m_N$ = 41.47 MeV fm$^2$, $m_N$=938.94 MeV with $\hbar$=197.32698 MeVfm \cite{data}.}
\end{table}
The MT-I-III(M) potential, which describes the $nn$- and $np$-components of the nucleon-nucleon interaction, leads to a significant change in the binding energy on the value of 160~keV. Note the paradoxical similarity this result with the result of realistic calculations from Ref. \cite{M21}, where the mass corrected $AAA$ model was used. The effect of charge symmetry breaking (CSB) on the $Nd$ breakup  was studied  within this phenomenological $AAB$ model in Ref. \cite{FSV19}.

However, in this work, we would like to focus only on the effect of the difference in nucleon masses.
In Tabl. \ref{t33}, the contributions of two terms of the Hamiltonian are listed with the notations: $H_0$ is the kinetic energy and $V_{2bf}$ is the energy of pair potentials.
The Hamiltonian is represented by the form: $\beta H_0 +\alpha V_{2bf}$, where $\beta$, $\alpha$ $\sim$ 1. For the Faddeev equation (\ref{eq:3}), we perform corresponding transformations:
\begin{equation}
H^a_0 \to \beta H^a_0, \quad a=U,W, \qquad V_b\to \alpha V_b, \quad b=nn,np. 
\end{equation}
Here, $H^a_0$ is transformed by changing the masses as $m^a \to m^a/\beta$.
Eq. (\ref{scal}) represents the  binding  $E(\beta, \alpha)$ is a sum of the terms 
$\langle H_0 \rangle$ and $\langle V_{2bf}\rangle$.
The contribution of each term of Hamiltonian is evaluated as
\begin{equation}
\langle H_0 \rangle=dE(\beta)/d\beta, \quad \langle V_{2bf}\rangle=dE(\alpha)/d\alpha
\label{der}
\end{equation}
in small vicinity of the point $\beta$=$\alpha$=1.
We evaluated the derivatives (\ref{der}) numerically using finite difference approximation. Near this point, the functions are almost linear, which ensures good accuracy of the numerical approximation. The functions $E(\beta, \alpha=1)$ and
$E(\beta=1, \alpha)$ are shown in Fig. \ref{fig:41a} in the range from 0.8 to 1.15. The obtained values are well approximated by cubic polynomials.
\begin{figure}[t]
\begin{center}
\includegraphics[width=16pc]{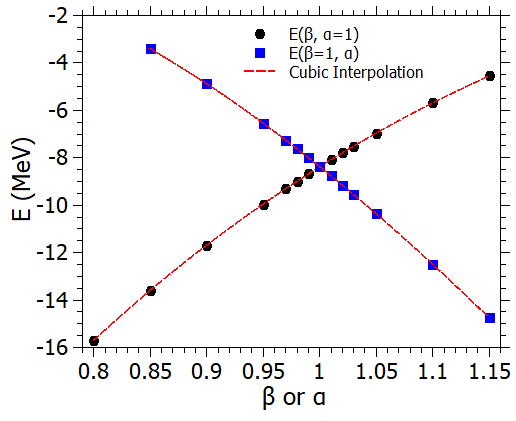}
\end{center}
\caption{The $^3$H ground state energy  $E$ as a function of scaling parameters $\beta$ and $\alpha$ within the $AAB$ model with MT-I-II(M) potential. The calculated values for $E(\beta, \alpha=1)$ and $E(\beta=1, \alpha)$ are shown by circles and rectangles, respectively.
The dashed lines correspond to the cubic polynomial  interpolation.
} \label{fig:41a}
\end{figure}

Taking into account the results of the estimation of $\langle H_0 \rangle$ and $\langle V_{2bf}\rangle$ given in Table. \ref{t33} we can conclude that the replacement of the $AAA$ model by the $AAB$ model leads to a decrease in the ground state energy by 0.2 MeV and a redistribution of the average contribution of the kinetic and potential terms of the Hamiltonian by 0.3 and 0.4 MeV. The corresponding energy change associated with the difference in nucleon masses gives 4.7~keV, as we have seen.
This result agrees with 6~keV realistic calculations made in Ref. \cite{M21}. The accuracy of our three-body energy calculations is estimated at four decimal places. \cite{M21} calculations are given with an accuracy of three decimal places.
Because of the small mass difference, these results appear to be the same. However, the difference between the two results is about 50\%. The mass difference effect is estimated in \cite{M21} as a matrix element of the correction of the kinetic operator. This correction is artificial for the $AAA$ model. The authors presented the results of calculations for the matrix element in keV with an accuracy of three decimal places. However, the estimate of the mass difference effect is given in \cite{M21} within the accuracy specified for the bound state energy according to Eq. (\ref{dm1}). This dualism of accuracy reflects the conceptual problem of the $AAA$ model mentioned above. Also, our result for the $^3$He mass difference effect disagrees with one obtained in \cite{M21} within the corrected $AAA$ model. We estimated the ground state energy of
-7.662~MeV for the case $ m_n$=$m_p$=$m_N$ and -7.8747~MeV for the case $ m_n\neq m_p$. It gives the mass effect for energy $\Delta E_m$=-7.4~keV that has larger on 20\% than the result -6~keV of \cite{M21}.

\section{ The \texorpdfstring{$AAA$}\ \ model: mass-energy compensation and nucleon effective mass}
\label{sec:AV14}
The $AAA$ model for three-nucleon bound states is applied to the study of nucleon-nucleon interactions based on the identity of protons and neutrons in isospin formalism when ignoring the difference of masses. One can find many examples of corresponding calculations in the literature.
In this section, we present the results of our calculations for the charge independent AV14 potential \cite{AV14}. For this potential, several papers of other authors are available \cite{C85,Nogga,KVR}. Among them, there are calculations with three-body potentials taken into account.
Our results for 3$N$ binding energies are presented in Table \ref{t1} together with those of other authors. In these calculations, a three-body potential was not applied.
\begin{table}[!bt]
\caption{
3$N$ binding energies $E^{Cal.}(m_N,m_N)$ calculated for AV14 $NN$ potential together with experimental values. The experimental data are from \cite{tunl1}. The energies are given in MeV.
}
\label{t1}
\begin{tabular}{lcccccccc}
\hline
{Pot.} &$J_{max}$ & $N_c$ &
\multicolumn{3}{c}{$^3$H} &
\multicolumn{3}{c}{$^3$He} \\
\noalign{\smallskip}\hline\noalign{\smallskip}
& & &\cite{Nogga} &\cite{KVR}&            & \cite{KVR}&\cite{C85} &  \\ \noalign{\smallskip}\hline\noalign{\smallskip}
AV14 & 2 & 18& -7.577 & -7.660&-7.578 & -7.010 & -6.920 &-6.913 \\
& 3& 26&-7.659& --&-7.661 & -- & -7011& -6.995 \\
& 4 &34& -7.674& -7.678 & -7.675 & -7.027&-7.014 &-7.009\\
& 5 &42&-7.680 &-- & -7.682 & -- & -- & -7.017 \\
& 6 &48&-7.682& -7.683 & -7.684 &-7.032 & -- &-7.018 \\
& 7 &55&-- & -- & -7.684 & -- & -- &-7.018 \\ \noalign{\smallskip}\hline\noalign{\smallskip}
Exp. \cite{tunl}& &&-8.481798 $\pm$ 0.000002  && &-7.718043 ± 0.000002 & &\\
\hline
\end{tabular}\\
\end{table}
The averaged nucleon mass is chosen as in Ref. \cite{Nogga} $m_N$=938.9 MeV (or $\hbar ^2/m_N$ = 41.473 MeV fm$^2$).
Taking into account the convergence of the calculation in $J_{max}$, we conclude that the numerical error of these calculations is less than 1~keV.
This is consistent with previous calculations by other authors.
A general description of the numerical methods used by us can be found in \cite{FM} and \cite{Su}.

Without the three-body potential, the 3$N$ system is weaker coupled, and the difference between our results and the experimental values
$\Delta E=E^{Cal.}(m_N,m_N)-E_{exp.}$ is 0.798~MeV for $^3$H and 0.701~MeV for $^3$He, respectively. Note that the channel $T$ = $\frac32$ is
ignored in these calculations.

We defined two notations: the effective mass of a nucleon in a nucleus is $m^*_N$, and the averaged free-nucleon mass is $m_N$. The effective mass is obtained by a correction for the averaged free nucleon mass: $m^*_N=m_N+\Delta m$.
\begin{table}[!ht]
\caption{3$N$ ground state energies $E$($^3$H) and $E$($^3$He) calculated using the AV14 potential and correction for mass $m_N+\Delta m$ (effective mass model). Here $\Delta m$=16.1 MeV.
   For comparison, the AV14 + TM(3BF) results from Refs. \cite{Nogga,KKF} and correspondence to the experimental data $E_{exp.}$ are shown. The energies are given in MeV.}
\label{t2}
{\begin{tabular}{lcccc} \hline\noalign{\smallskip}
Model &  $E$($^3$H) & $E$ ($^3$He) & $(E-E_{exp.})$($^3$H) & $(E-E_{exp.})$($^3$He)\\ 
  \noalign{\smallskip}\hline\noalign{\smallskip}
Effective mass & -8.475 & -7.794 & 0.007 & -0.076 \\
3BF (TM)\cite{Nogga}$^*$ & -8.475 &-- & 0.007 & -- \\
3BF (TM)\cite{KKF} & -8.430 & -7.757 & 0.052 & -0.039 \\
\noalign{\smallskip}\hline
\end{tabular}\\
\smallskip
\noindent $^*$ 3NF calculated with jmax = 5 for inner basis. \ \ \ \ \ \ \ \ \ \ \ \ \ \ \ \ \ \ \ \ \ \ \ \ \  \ \ \ \ \ \ \ } 
\end{table}
\begin{table}[!ht]
\caption{
The $^3$H binding energies for AV14, AV14+ effective mass, and AV14 +TM(3BF) models, and the contributions of each term of Hamiltonian are shown in MeV. }
\label{t3}
\begin{center}
{\begin{tabular}{cccccc} \hline\noalign{\smallskip}
Model &$J_{max}$ & $E$($^3$H) & $\langle H_0 \rangle$ & $\langle V_{2bf}\rangle$ & $\langle V_{3bf}\rangle$ \\ \noalign{\smallskip}\hline\noalign{\smallskip}
AV14 \cite{Nogga}&6 & -7.682 & 45.680 & -53.364 & -- \\
AV14 \cite{KVR}&6 & -7.683 & 45.767 & -- & -- \\
AV14 &7 & -7.684 & 45.7 & -53.4 & -- \\ \noalign{\smallskip}\hline\noalign{\smallskip}
AV14 +TM(3BF) \cite{Nogga}&5 & -8.475 & 49.321 & -56.498 & -1.300 \\
&6 & -8.486 & 49.340 & -56.518 & -1.310 \\
AV14 +TM(3BF) \cite{KVR}& 6 & -8.480 & 49.30& -- & -- \\
AV14 +effective mass &7 & -8.475 & 47.4 & -55.9 & -- \\
& & & (49.3-1.9) & (-56.5+0.6) & \\
\noalign{\smallskip}\hline
\end{tabular} }
\end{center}
\end{table}
In Tabl.~\ref{t2}, we show the results of our $AAA$ calculations with $\Delta m$=16.1~MeV and with using the AV14 potential for the bound state energy of $^3$H and $^3$He nucleus. The value $\Delta m$ was adjusted to reproduce the experimental data. The results were compared to the ones from Refs. \cite{Nogga} and
\cite{KKF} where the AV14 and TM (three-body force) potentials were applied. One sees that our effective mass model leads to similar results, $E^{Cal.}_t(m_N,m_N)+\Delta E^{3NF}\approx E^{Cal.}_t(m^*_N,m^*_N)\approx E_{exp.}$. At the same time, the experimental $^3$H ($^3$He) energy is slightly under (over) estimated in all presented calculations. We should note that the experimental values are only used as reference points for these calculations.

\begin{figure}
\begin{center}
\caption{ The $^3$H and $^3$He ground state energy $E$ as a function of the nucleon effective mass $m_N^*/m_N=1/\beta$.
Open circles are corresponding to the calculated value for $^3$H nucleus. The filled  circles are the results for $^3$He nucleus.
 The dashed (solid) fine lines show correspondence between the $^3$H energy and the effective mass of nucleon $m_N^*/m_N$ for $m_N^*/m_N=1$ and 
 $m_N^*/m_N$ about 1.017 and the experimental value for the ground state energy about -8.48 MeV.
} \label{fig:3a}
\includegraphics[width=18pc]{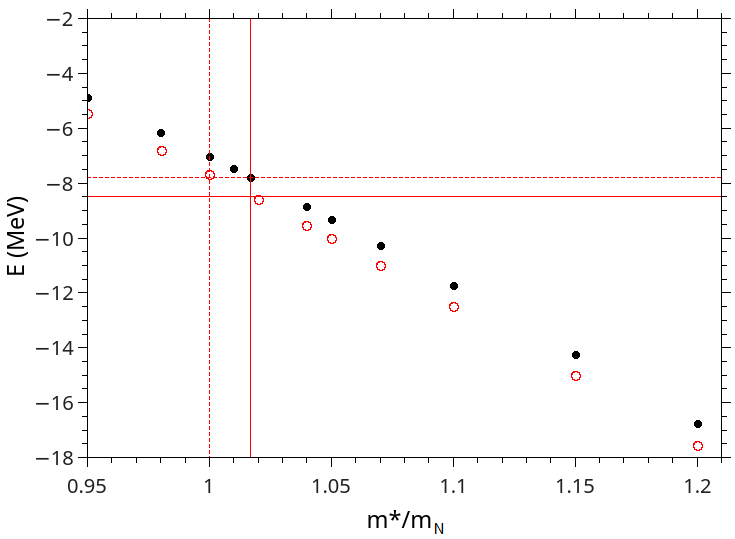}
\end{center}
\end{figure}
The procedure of the adjustment for the nucleon mass is presented in Fig. \ref{fig:3a}  for $^3$H. The mass scaling produces binding energy values around the experimental point of -8.48~MeV.
We show that the linear function can describe the energy-mass dependence approximately and establishes the correspondence between the experimental value of the energy and nucleon effective mass. 

The $2N$ bound problem can also exhibit such linear property. Our calculations for deuteron performed with the AV14 potential within the same mass ranges show the well-known dependence of the binding energy on the particle mass like $E \sim \frac1{m}$, which gives a linear approximation within only small vicinity of the point $m^*/m_N=1$. In this point,  the kinetic energy of the deuteron has been estimated as 19.2~MeV. The results of our calculations for the deuteron binding energy 
are shown in Fig. \ref{fig:5s}  for the effective mas  variations from 0.9$m_N$ to 1.3$m_N$. We assumed that  the value $m^*/m_N-1$ is in the range $0 < m^*/m_N -1 <1$. 
\begin{figure}[t]
\begin{center}
\includegraphics[width=18pc]{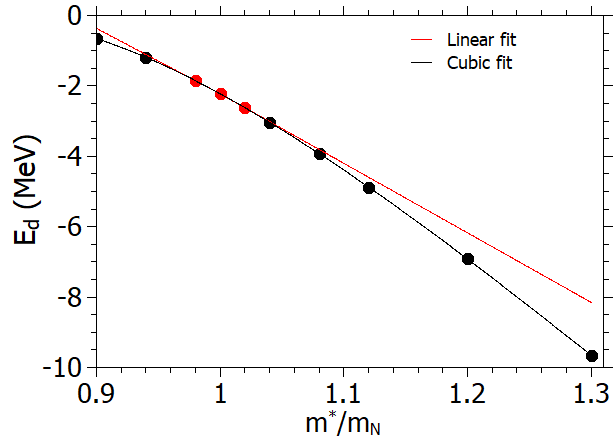}
\end{center}
\caption{The ground state energy $E_d$ of the deuteron as a function of the effective mass calculated with the AV14 nucleon-nucleon potential. The black curve corresponds to the cubic fit for the energy-mass dependence.
 The red dashed line gives the linear fit for the calculated energies shown by red symbols. 
} \label{fig:5s}
\end{figure}
In Fig. \ref{fig:5s} we present two fittings for the calculated results.
A linear function can be a good approximation for the function $1/{m^*}$ in the vicinity $\pm0.03 m_N$ of the point $m^*/m_N=1$. When the vicinity is expanded by a factor of ten, a good approximation by a cubic function was obtained. In this way, we take into account the first three terms of the the Taylor series for the function $1/{m}$. The energy can be written as $E(m^*)=E(m^*=m_N)(1-\sum_{n=1,2,...}(-1)^n(m^*/m_N-1)^n)$.

From Figs.  \ref{fig:3a} and \ref{fig:5s} one can see that mass scaling gives  different  $3N$ and $2N$ energy-mass dependences. The resulting approximations are linear and cubic, respectively, in the same range of mass changes.
This result is interpreted as  the scaling $f(\xi x)$ for a function $f(x)$. The parameter $\xi$ for the $3N$ system determinate by
the ratio      $\langle H_0 \rangle_d/\langle H_0 \rangle_t$, where the deuteron and the triton matrix elements are marked by the index $d$ or $t$, respectively.
These matrix elements differ by a factor of two.
\begin{figure}[t]
\begin{center}
\includegraphics[width=16pc]{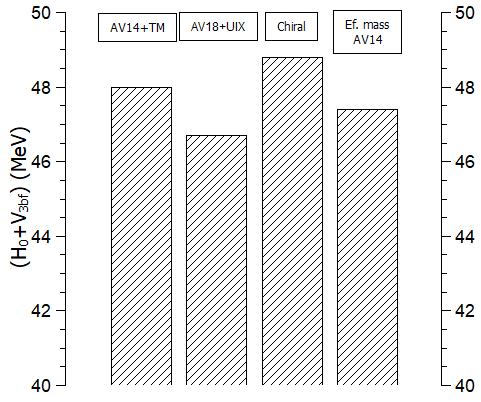}
\end{center}
\caption{The averaged values for $H_0+V_{3bf} $ evaluated with different inputs of the $^3$H bound state problem. The
results are from \cite{Nogga} (AV14+TM), \cite{pudiner} (AV18+UIX), \cite{Sk} (Chiral) and our eavaluation (Ef. mass AV14).
} \label{fig:4a}
\end{figure}

To evaluate the averaged contribution of the kinetic and potential energy terms of the Hamiltonian, we performed calculations, using the scaling of Eq. (\ref{scal}): $\beta H_0+\alpha V$, where $\beta, \alpha> 0$, and
where $m \to m/\beta$ in the $H_0$.
Averaging, we have $E=\beta \langle H_0 \rangle+ \alpha \langle V \rangle$ and $\langle V \rangle=dE(\alpha)/d\alpha$ and $\langle H_0 \rangle=dE(\beta)/d\beta$ calculated for
$\alpha=\beta =1$ according to Eq. (\ref{der}).
The results for the $^3$H binding energy are shown in Tabl. \ref{t3} along with the results from Refs. \cite{Nogga,KVR}.
The calculations have been performed with and without using the three-body force. In the case, when the three-body potential has been ignored, the results of other authors are in good agreement with ours.
The three-body potential may be considered as a perturbation contributing with -1.3~MeV to the potential energy term $\langle V_{2bf}\rangle$=-56.518~MeV. This evaluation has been obtained in Ref. \cite{Nogga}. Also, the calculation presented in Ref. \cite{Sk} shows similar values: $\langle H_0 \rangle$=48.088~MeV, $\langle V_{2bf}\rangle$=-55.187~MeV, $\langle V_{3bf}\rangle$=-1.381 MeV, obtained with a chiral N$^3$LO potential. We can note that the $^3$H results obtained in Ref. \cite{pudiner} using the combination AV18+UrbanaIX(3NF) for interactions are also comparable with ones for the set AV14+TM(3NF): $\langle H_0 \rangle$=50.0~MeV, $\langle V_{2bf}\rangle$=-58.24~MeV, $\langle V_{3bf}\rangle$=-1.20~MeV.
For comparison, these $^3$H data are presented by the diagrams in Fig. \ref{fig:4a}. We compare the averaged values obtained for the matrix element $\langle H_0+V_{3bf}\rangle$. One can see that
our result does not go beyond the range of variations of these values of different models.

The difference in kinetic energy terms of our result and the results of Refs. \cite{Nogga,KVR} is about 1.9 MeV, and the deference between potential energy terms is about  -0.6~MeV (given in Tabl. \ref{t3} in parentheses). Thus, we obtain the 1.3~MeV for the compensation contribution of kinetic and potential energy for the 3BF term found in Ref. \cite{Nogga} with the value of -1.3 MeV. This compensation effect can be interpreted as a manifestation of the equivalence of mass and energy in three-nucleon systems.  Also, in Fig.  \ref{fig:3a}, the dependence of energy on mass is linear like $\Delta E \approx Const \Delta m^*$ in a wide range near the point $m^*/m_N=1$. One can see an analogy with the well-known formula $E=mc^2$, which connects energy and mass.
 
It should be noted that the effective mass does not attribute to free particles or free-bound pairs of particles.
This consideration is limited by the deeply bound states, in which the three-body effects are significant. It is assumed that a two-body cluster (for example, a deuteron)
existing in a three-body system can be distorted. Namely, the binding energy of a particle pair should be changed, because of the mass adjustment.
The effective masses for bound excited states may 
depend on binding energy measured from the nearest two- or three-body threshold.
An analogy for such energy-dependent effective mass could be found in solid-state physics \cite{FSV06} or in many-body physics \cite{GSV}.
The effective-mass approach is a phenomenological technique to describe bound systems with three-body interactions based on the mass-energy compensation effect.
We have shown that the three-body energy is a linear function of the parameters $\alpha$ and $\beta$ near the point $\alpha$=$\beta$=1. The factors $\alpha$ and $\beta$ change the energy with opposite signs and make the mass-energy compensation possible.
The effective mass depends on the interaction proposed for such compensation. For example,
in the case of repulsive three-body force, the effective mass could be smaller than the free mass.
\begin{figure}[!ht]
\begin{center}
\includegraphics[width=18pc]{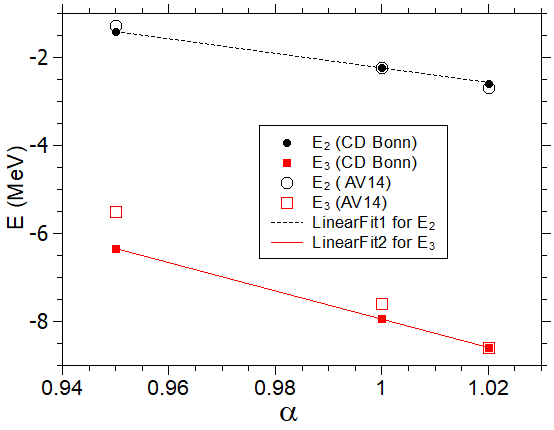}
\end{center}
\caption{The dependence of two and three nucleon binding energies ($E_2$) and  ($E_3$) on the scaling parameter $\alpha$   for the CD Bonn and AV14 potentials near the point $\alpha$=1.  The results of the calculations \cite{Witala} are marked by solid symbols. Our results are marked by the open symbols. The linear fits (LinearFit1 and LinearFit2 for $E_2$ and $E_3$, respectively) for the CD Bonn results are shown.
} \label{fig:6}
\end{figure}
It would be interesting to analyze the result of the recent calculations \cite{Witala} as an example of the compensation effect. In Ref. \cite{Witala}, the $s$-wave component of CD Bonn $NN$ potential was scaled by the factor $\alpha$ to study the effect of this adjustment on fitting the $nd$-scattering experimental data. Such scaling can be equivalent to a correction of the nucleon mass, as we have shown above. 
The two and three-body binding energies have been increased/decreased from the values corresponding to the original CD Bonn potential and $\alpha$=1. When $\alpha$=1.02, the values -2.592~MeV and -8.598~MeV were calculated for $^2$H and $^3$H, respectively. At the same time, the two-body binding energies calculated with these potentials are close to the experimental value of -2.224575(9)~MeV \cite{2H}.
Our calculations give similar values taking into account the different types of pair potentials and the difference in scaled partial components. 
Our results are -2.650~MeV and -8.475~MeV when $\beta$=0.98. 
The original values (when $\alpha$ and $\beta$ are equal to 1)  for three-body binding energy with CD Bonn and AV14 potentials are -7.923~MeV and -7.684~MeV, respectively. 
In Fig. \ref{fig:6}, we present the linear dependence of the energies on the scaling parameter $\alpha$ near the point $\alpha$=1 obtained by using the results from Ref. \cite{Witala} for CD Bonn potential. This relationship is consistent with our mass scaling results shown in Fig. \ref{fig:6} for comparison.
Another example of the mass-energy compensation effect described by Eq. (\ref{scal}), one can find in Ref. \cite{F90} where a dependence of the matrix element of the kinetic operator $\langle H_0 \rangle $ on the binding energy has shown for $^3$H. 
This dependence corresponds to a linear function described by Eq. (\ref{scal}).

\section{Summary}The defect mass formula reflects the fundamental equivalence of mass and energy in a bound nuclear system.
According to this formula, including experimental values for neutrons and protons masses, one must consider the $^3$H and $^3$He nuclei as $AAB$ systems.
The AAA model with averaged nucleon mass is widely used for realistic three-nucleon calculations in which the bound state energy and other observable quantities are the targets. However, formally applying the defect mass formula within the $AAA$ model generates two conceptual problems. This problem can be solved by using the $AAB$ model.

Namely,  the defect mass formula leads to the difference between the $AAA$ and $AAB$ models which is about 0.6~MeV, which is large and comparable with the energy shift between the $^3$H and $^3$He ground state.
Another problem relates to the cyclic permutation symmetry of the $AAA$ system. The physical $AAB$ system breaks this symmetry to reduce it to the exchange symmetry for the pair of identical particles.
From the point of view of rigorous $3N$ calculations, one has to consider the $AAA$ model to be approximate.
 As an example of an $AAB$ treatment, we presented a model based on phenomenological $nn$, $np$, and $pp$ potentials.
  We evaluated directly the energy shift induced by the mass difference with accuracy up to 0.1~MeV. The comparison with the similar evaluation for the $AAA$ model \cite{M21}   shows a visible difference at least in 1.3~keV (1~keV for $^3$He). These results can be explained by a difference in used potentials.
  
The problem of equal particle masses of the $AAA$ model can be formally solved by using realistic masses in an approximative procedure \cite{K2008} that modifies the kinetic-energy operator perturbatively by using an isospin-dependent mass correction.
This correction does not convert $AAA$ equations to $AAB$ ones and does not change the cyclic permutation symmetry of the wave function which still corresponds to the $AAA$ model. 
Considering the $AAA$ model, we proposed modified mass correction for the $AAA$ model. This approach is isospin independent in contrast to the one from \cite{K2008}. The results for the binding energy obtained within the framework of these various correction procedures may be different, which can be interpreted as the uncertainty of such $AAA$ calculations.

Finally, we would note that the disagreement between the $AAA$ model and the fundamental physical relation is an important reason for the future development of rigorous three-body calculations.  However, the price for $AAB$ calculations is high because one needs to make a significant reformulating of the $3N$ formalism and correction of computer codes created over the past decade.

Our $AAA$ calculation for $^3$H ($^3$He) nucleus has demonstrated another manifestation of the mass and energy equivalence. Utilizing the results of different authors, we have shown that nucleon-mass correction would be approximately equivalent to the energy of attractive contribution of three-nucleon potential. To clarify this result, we have considered the mechanism of mass-energy compensation, which allows us to introduce the effective mass of nucleon in a bound three-nucleon system. In the linear combination $E(\beta,\alpha)=\beta \langle H_0\rangle+\alpha \langle V \rangle$, the kinetic and potential terms have opposite signs. A variation of the $\beta$ parameter can be compensated by a variation of the $\alpha$ parameters.
Locally, for small variations of the $beta$ ($\alpha$) near the point $\beta=1$ ($\alpha=1$), the energy dependence on $\beta$ ($\alpha$) is close to being linear. That allows us to apply the numerical estimation $\langle H_0 \rangle=dE(\beta)/d\beta$ and $\langle V \rangle=dE(\alpha)/d\alpha$ for the averaged kinetic and potential energies with acceptable accuracy.
Generally the function  $E(\beta,\alpha)$ is nonlinear.

However,  a linear dependence of the energy on the effective mass $m^*/m_0=1/\beta$ was found
in the region from 0.9~$m_0$ to 1.2~$m_0$.
In our approach, an energy shift corresponds to the small variations of $m^*/m_0$  by the formula $$\Delta E\approx -\langle H_0\rangle \Delta (m^*/m_0)$$  near the point $m^*/m_0$=1.  This relation, written as $dE/d(m^*)=Const$, connects variations of energy and mass similar to the well-known formula $E=mc^2$. The large value of the kinetic energy term ($\langle H_0\rangle \approx $ 45~MeV, with using the AV14 potential) provides a large binding energy shift induced by a small mass correction. For the effective mass found in our study with the value of $m^*/m_0$=1.02, the binding energy shift $\Delta E$ can be roughly estimated at about -0.9~MeV because $\Delta E \approx -$ 45~MeV 0.02=-0.9~MeV. This value is related to the contribution of the three-body force term (like the TM potential).  We estimate that a 2\% mass correction gives the response of 12\% in the binding energy, in scaled energy and mass values.
It can be expected that the subtle details of the three-body potential will have a distinct effect on the three-body energy within this 12\% correction. However, it is possible to consider an effective representation of this potential using the effective mass method within the framework of the phenomenological approximation for 3$N$ systems, when the task of studying the three-body force is not at stake.
In this regard, it can be noted that the effective mass method can be applied to three-particle cluster models that include three-body interaction. In Ref. \cite{FKV22}, this method was used for a study of the $^{12}$C nucleus within the 3$\alpha$-particle model \cite {FSV05, Mo}.  

\section*{Acknowledgments}
This work is supported by US National Science Foundation HRD-1345219 award and the Department of Energy/National Nuclear Security Administration award NA0003979.


\end{document}